\documentclass[showpacs,preprintnumbers,amsmath,amssymb]{revtex4}
\usepackage{wrapfig}
\usepackage{graphicx}
\usepackage{dcolumn}
\usepackage{bm}

\begin{document}


\title{Stochastic dynamics of N bistable elements with global time-delayed interactions: towards an exact solution of the master equations for finite N}
\author{M. Kimizuka and T. Munakata}
\affiliation{Department of Applied Mathematics and Physics, 
Graduate School of Informatics, Kyoto University, Kyoto 606-8501, Japan
} 
\author{M. L. Rosinberg}
\affiliation{Laboratoire  de  Physique  Th\'eorique  de  la  Mati\`ere
Condens\'ee,  CNRS-UMR 7600,  Universit\'e  Pierre et  Marie Curie,  4
place Jussieu, 75252 Paris Cedex 05, France} 

\begin{abstract}
We consider a network of $N$ noisy bistable elements with global time-delayed couplings. In a two-state description, where elements are represented by Ising spins, the collective dynamics is described by an infinite hierarchy of coupled master equations which was solved at the mean-field level in the thermodynamic limit.  For a finite number of elements, an analytical description was deemed so far intractable and numerical studies seemed to be necessary. In this paper we consider the case of two interacting elements and show that a partial analytical description of the stationary state is possible if the stochastic process is time-symmetric. This requires some relationship between the transition rates to be satisfied.
\end{abstract}
\pacs{02.50.Ey, 05.40.-a, 05.10.Gg}
\maketitle

\section{Introduction\label{sec:1}}

Systems with time-delayed interactions have been a subject of extensive studies in recent years due to their relevance to a wide range of phenomena occurring in physics, biology, ecology, economics, and other sciences. 
In most situations,  the effect of random noise due to the environmental fluctuations cannot be ignored and observables  are better described as stochastic variables. As is well known, this may have a major impact on the dynamical behavior, in particular when noise combines with nonlinearity, which leads to many remarkable effects such as dynamical transitions\cite{BPT1994},  synchronization\cite{PRB2003}, stochastic resonance\cite{GHJM1998}, coherence resonance\cite{PK1997}, etc... The addition of time delay increases the dimensionality and hence the complexity of these systems, inducing new phenomena such as multi-stability\cite{YS1999} and oscillatory behavior\cite{AB,BVTH2005,CWZA2005}.

By definition, stochastic time-delay systems are non-Markovian, which seriously complicates the analytical treatment as the standard tools for ordinary stochastic differential equations are not directly applicable. Although one can extend the Fokker-Planck description to stochastic delay-differential equations\cite{GLL1999,F2002}, exact solutions are rare (essentially limited to the linear case\cite{KM1992,GLL1999,FB2001}), and in order to calculate probability densities and time correlation functions one has to resort to approximate treatments ({\it e.g.} small-delay expansion\cite{GLL1999,ZXXL2009}, perturbation theory\cite{F2005}) and more generally to numerical simulations. This can be traced back to the fact that stochastic delay systems can be viewed as systems with an infinite number of degrees of freedom. 

The situation is even more complicated when one considers several interacting units with time-delayed couplings, a situation that usually occurs in a biological context ({\it e.g.} in neural or genetic regulatory networks) and can also be realized with laser networks (see {\it e.g.} \cite{HGBMH2004,GMTG2007} for recent references). From a theoretical point of view, the canonical example is a globally coupled network of stochastically driven bistable elements, a system that has been extensively studied in the absence of delay\cite{DZ1978,JBPM1992,GHJM1998}.  As is often done with bistable elements, one may replace the original continuous system by a two-state model with suitably chosen transition rates and replace the stochastic differential equations by master equations for the occupation probabilities. Then one has to cope with an infinite hierarchy of coupled equations which at first sight cannot be closed. So far, a full analytical description of the dynamics is only avail
 able when there is a single element\cite{TP2001,M2003} or an infinite number of elements\cite{HT2003}. In this thermodynamic limit, which may be justified in actual situations ({\it e.g.} a multicellular system\cite{CWZA2005}), one can derive a deterministic equation of motion for the mean-field variable {\it i.e.} the ensemble average of the state variable. The solution then exhibits phase transitions to nontrivial stationary states or delay-dependent oscillations via Hopf bifurcations. In principle, corrections to the mean-field behavior can be obtained within an expansion in the inverse system size (see {\it e.g.} \cite{G2009}).

The aim of the present work is to show that a solution of the delay master equations can also be obtained for a {\it finite} number of interacting bistable elements, at least in the stationary state and for a restricted time interval. In the following, for simplicity, we only treat the case of two coupled elements but the demonstration can be extended to several units at the price of increasing analytical complexity. Interestingly, the demonstration used the time-symmetry of the delay master equations, an issue which does not seem to have been discussed in the existing literature. 

The paper is organized as follows.  In the next Section we  present the model and derive the master equations for the probabilities. In Section III we solve these equations and compute the time correlation functions under the condition that a certain relation between the transition rates is satisfied. Analytical results are then compared to the results of numerical simulations. Concluding remarks are given in Section IV.
Appendix A is devoted to the  analysis of time-symmetry and Appendix B details the solution of the master equations.

\section{Model and master equations}

As Huber and Tsimring\cite{HT2003} we consider an ensemble of $N$ identical bistable elements, each of them characterized by the variable $x_i(t)$ and obeying the coupled Langevin equations
\begin{equation}
\label{Eq1}
\dot{x}_i(t)=-\frac{dV(x_i)}{dx_i}+\epsilon X (t-\tau)+\sqrt{2D}\xi_i(t) \ , \ i=1...N
\end{equation}
where $V(x)=-x^2/2+x^4/4$ is a generic symmetric double-well potential and $X(t)$ is the global `mean-field' 
\begin{equation}
X(t)=\frac{1}{N}\sum_i x_i(t) \ .
\end{equation}
Here $\tau$ is the time delay, $\epsilon$ is the strength of the feedback coupling, and $D$ is the variance of the Gaussian fluctuations, which are $\delta$-correlated and mutually independent $<\xi_i(t)\xi_j(t')>=\delta(t-t')\delta_{ij} $. 
For each $i$, this set of equations thus describes the overdamped motion of a particle evolving in an effective  $\tau$-dependent double-well potential $U_{\tau}(x)=V(x)-\epsilon X(t-\tau)x$. Note that each element is identically coupled to {\it all} units at time $t-\tau$, including itself (the case of a chain with  unidirectional coupling, {\it i.e.} $x_i(t)$ only coupled to $x_{i-1}(t-\tau)$, is much simpler and has been considered in Ref.\cite{KM2009}).

In the following we shall be interested in the stationary state that is reached in the large-time limit. Since the number of elements is strictly finite, one expects this state to be unique for all couplings with an average value  $<X(t)>=0$ ({\it i.e.} there is no phase transition).

As in Refs.\cite{TP2001,HT2003} we consider the case of small noise and small coupling where one can neglect the intrawell fluctuations and replace the continuous dynamical variables $x_i(t)$ by the two-state variables $s_i(t)$  that take the values $\pm 1$. The switching rates associated to the instantaneous potential $U_{\tau}(x)$ can be calculated from Kramers formula\cite{K1940} $\gamma_K=(2\pi)^{-1}\sqrt{U''_{\tau}(x_{\pm})U''_{\tau}(x_0)}\exp(-\Delta U_{\tau}/D)$ where $x_{\pm}$ and $x_0$ are the positions of the minima and the maximum of the potential, respectively, and $\Delta U_{\tau}=U_{\tau}(x_0)-U_{\tau}(x_{\pm})$ is the barrier that an element has to overcome to jump from one stable state to the other. For small  $\epsilon$, the two minima of the potential  are located at $x_{\pm}=\pm 1+\epsilon X/2$, which yields\cite{HT2003}
 \begin{equation}
 \label{Kramers}
\gamma_K=\frac{\sqrt{2\pm 3\epsilon X(t-\tau)}}{2\pi} \exp \big(-\frac{1\pm 4\epsilon X(t-\tau)}{4D}\big) \ ,
\end{equation}
so that $\gamma_K$ can take different values depending on the sign of $s_i$ at time $t$ and the state of the system at time $t-\tau$ (in the two-state description $X(t)=(1/N) \sum_i s_i(t)$). When $N=1$, which is the simplest case studied by Tsimring and Pikovsky\cite{TP2001}, there are only two possible values 
\begin{eqnarray}
\label{RatesN1}
\gamma_{1,2} &=&\frac{\sqrt{2\pm3\epsilon}}{2\pi} \exp\big(-\frac{1\pm4\epsilon}{4D}\big)  
\end{eqnarray}
corresponding  to $s(t)s(t-\tau)=1$ and $s(t)s(t-\tau)=-1$, respectively. In the opposite limit studied in Ref.\cite{HT2003}, where the number of units is very large,  stochastic fluctuations can be neglected and the collective variable $X(t)$ approaches the average value $<s(t)>=\sum_{s=\pm 1} s p(s,t)$, where $p(\pm 1,t)$ are the occupation probabilities of the states $s=\pm 1$. One can then derive a closed equation of motion for the mean field $X(t)$.  

In the present work we focus on the case $N=2$ and from Eq. (\ref{Kramers}) we must take into account three different switching rates
\begin{eqnarray}
\label{RatesN2}
\gamma_0 &=&\frac{\sqrt{2+3\epsilon}}{2\pi} \exp\big(-\frac{1+4\epsilon}{4D}\big)  \ \ \mbox{for} \ \ \vert X(t-\tau)\vert=1 \  \mbox{and} \  s_i(t)X(t-\tau)=1 \nonumber\\
\gamma_1 &=&\frac{\sqrt{2}}{2\pi} \exp\big(-\frac{1}{4D}\big)\ \ \mbox{for}  \ X(t-\tau)=0\nonumber\\
\gamma_2 &=&\frac{\sqrt{2-3\epsilon}}{2\pi} \exp\big(-\frac{1-4\epsilon }{4D}\big) \ \ \mbox{for} \ \ \vert X(t-\tau)\vert=1  \ \mbox{and} \  s_i(t)X(t-\tau)=-1 
\end{eqnarray}
with $X(t)=(1/2)[s_1(t)+s_2(t)]$. These different rates can be put together in a single expression defining the switching rate from $s_i(t)$ to $-s_i(t)$ depending on the state of the spins $s_1$ and $s_2$ at $t-\tau$,
\begin{equation}
\label{Rates}
T(s_i\rightarrow -s_i;{\bf s})=\gamma_1+\frac{1}{4}(\gamma_0-\gamma_2)s_i(t)[s_1(t-\tau)+s_2(t-\tau)]+\frac{1}{4}(\gamma_0+\gamma_2-2\gamma_1)[1+s_1(t-\tau)s_2(t-\tau)] 
\end{equation}
where ${\bf s}$ denotes the two-component vector $\{s_1,s_2\}$ and it is implicit in the notation $T(s_i\rightarrow -s_i;{\bf s})$ that $s_i$ is taken at time $t$ and  ${\bf s}$ at time $t-\tau$.

The description of the dynamics of the system is then encoded in $4$ coupled master equations for the probabilities $p({\bf s},t)$ which read 
\begin{align}
\label{Master1}
\dot{p}(\{s_1,s_2\},t)= 
&\sum_{{\bf s}'} \Big[T(-s_1\rightarrow s_1;{\bf s}')p({\bf s}',t-\tau;\{-s_1,s_2\},t)+ T(-s_2\rightarrow s_2;{\bf s}')p({\bf s}',t-\tau;\{s_1,-s_2\},t) \nonumber\\
&-[T(s_1 \rightarrow -s_1;{\bf s}')+T(s_2\rightarrow -s_2;{\bf s}')]p({\bf s}',t-\tau;\{s_1,s_2\},t)\Big]
\end{align}
where $p({\bf s}' ,t-\tau;{\bf s},t)$ is the joint probability that the system is in state  ${\bf s}'$ at $t-\tau$ and ${\bf s}$ at $t$. The equations of motion are thus not closed at the level of one-time probabilities and one also need to consider the equations of motion for $p( {\bf s}' ,t-\tau;{\bf s},t)$, $p({\bf s}'',t-2\tau;{\bf s}' ,t-\tau;{\bf s},t)$, and so on. As already pointed out, this hierarchical structure merely reveals that a time-delayed system is a system with an infinite number of degrees of freedom.  For $N=1$, the master equations for the one-time probabilities $p(\pm 1,t)$ are closed\cite{TP2001} because one can use the exact relations $p(1,t-\tau;\pm 1,t)+p(-1,t-\tau;\pm1,t)=p(\pm 1,t)$ and  $p(\pm 1,t-\tau;1,t)+p(\pm 1,t-\tau;-1,t)=p(\pm 1,t-\tau)$ to eliminate the two-time probabilities. For similar reasons, a closed chain with unidirectional couplings can also be solved at the level of the one-time probabilities\cite{KM2009}. On the other 
 hand, for the system described by the coupled Langevin equations (\ref{Eq1}), one can easily check that there are not enough such exact relations to close the hierarchy for a generic value of $N$, even when taking into account the additional symmetry relations obtained by exchanging the spins $1$ and $2$ and the signs $+1$ and $-1$ (see below).

\section{Stationary solution of the master equations for $N=2$} 

We now turn our attention to the stationary solution of the $N=2$ model which satisfies $\dot{p}_{st}({\bf s},t)=0$. We are interested in calculating $p_{st}({\bf s})$ and the self and cross time-correlation functions $\psi_s(t)$ and $\psi_c(t)$
defined by 
\begin{align}
\psi_s(t)&=<s_1(t)s_1(0)>=\sum_{{\bf s}^0,{\bf s}} s_1s_1^0 \: p_{st}({\bf s},t|{\bf s}^0)\: p_{st}({\bf s}^0) \nonumber \\
\psi_c(t)&=<s_1(t)s_2(0)>=\sum_{{\bf s}^0,{\bf s}} s_1s_2^0\: p_{st}({\bf s},t|{\bf s}^0)\: p_{st}({\bf s}^0) 
\end{align}
where we have introduced the conditional probabilities $p_{st}({\bf s},t|{\bf s}^0)$ (hereafter it is implicit that ${\bf s}^0$ denotes the state at $t=0$). From Eqs. (\ref{Master1}), we readily get
\begin{align}
\label{Master11}
\dot{p}({\bf s},t|{\bf s}^0)=&\sum_{{\bf s}'} \Big[T(-s_1\rightarrow s_1;{\bf s}')p({\bf s}',t-\tau;\{-s_1,s_2\},t|{\bf s}^0)+ T(-s_2\rightarrow s_2;{\bf s}')p({\bf s}',t-\tau;\{s_1,-s_2\},t|{\bf s}^0)\nonumber\\
&-[T(s_1 \rightarrow -s_1;{\bf s}')+T(s_2\rightarrow -s_2;{\bf s}')]p({\bf s}',t-\tau;\{s_1,s_2\},t|{\bf s}^0)\Big] 
&
\end{align}
which, unsurprisingly, indicates that the calculation of $\psi_s(t)$ and $\psi_c(t)$ requires the knowledge of the three-time conditional probabilities $p_{st}({\bf s}',t-\tau;{\bf s},t|{\bf s}^0)$ which in turn depend on four-time functions, etc...

Hence, at this stage, it seems that the problem cannot be solved exactly. However, remarkably, an analytical  solution does exist if the switching rates satisfy the relation
\begin{align}
\label{Eqgamma}
\gamma_0\gamma_2=\gamma_1^2  
\end{align}
which may also be viewed as a natural consequence of Kramers' equations  (\ref{RatesN2}) if one  expands the product $\gamma_0\gamma_2$ up to order $\epsilon$.  Indeed, as shown in Appendix A, the stochastic process is then statistically time reversible in the stationary state. In other words, any sequence of states $S$ has the same probability as its time reverse $\overline {S}$:
\begin{align}
p_{st}(S) =p_{st}(\overline {S}) \ .
\end{align}
This readily implies that
\begin{equation}
p_{st}({\bf s}',t-\tau; {\bf s}^0,0; {\bf s},t)=p_{st}({\bf s},-t;  {\bf s}^0, 0; {\bf s}',\tau-t)
\end{equation}
and thus
\begin{equation}
\label{Timesym}
p_{st}({\bf s}',t-\tau;{\bf s},t\vert {\bf s}^0)=p_{st}({\bf s},-t;  {\bf s}',\tau-t\vert {\bf s}^0) \ .
\end{equation}
(In Appendix A we show an example where Eq.(\ref{Eqgamma}) is violated and thus Eq. (\ref{Timesym}) is not satisfied). From this equation we can derive a {\it closed} equation for the key quantity  $p_{st}({\bf s}',t-\tau;{\bf s},t|{\bf s}^0)$ and then compute the stationary occupation probabilities and the time correlation functions. We stress, however, that the solution for the time correlation functions is only valid in the interval $0\le t\le \tau$ (whereas Eq. (\ref{Timesym}) holds for all times $t$). In this sense the problem is only partially solved.
\begin{figure}[hbt]
\begin{center}
\includegraphics[width=9.8cm]{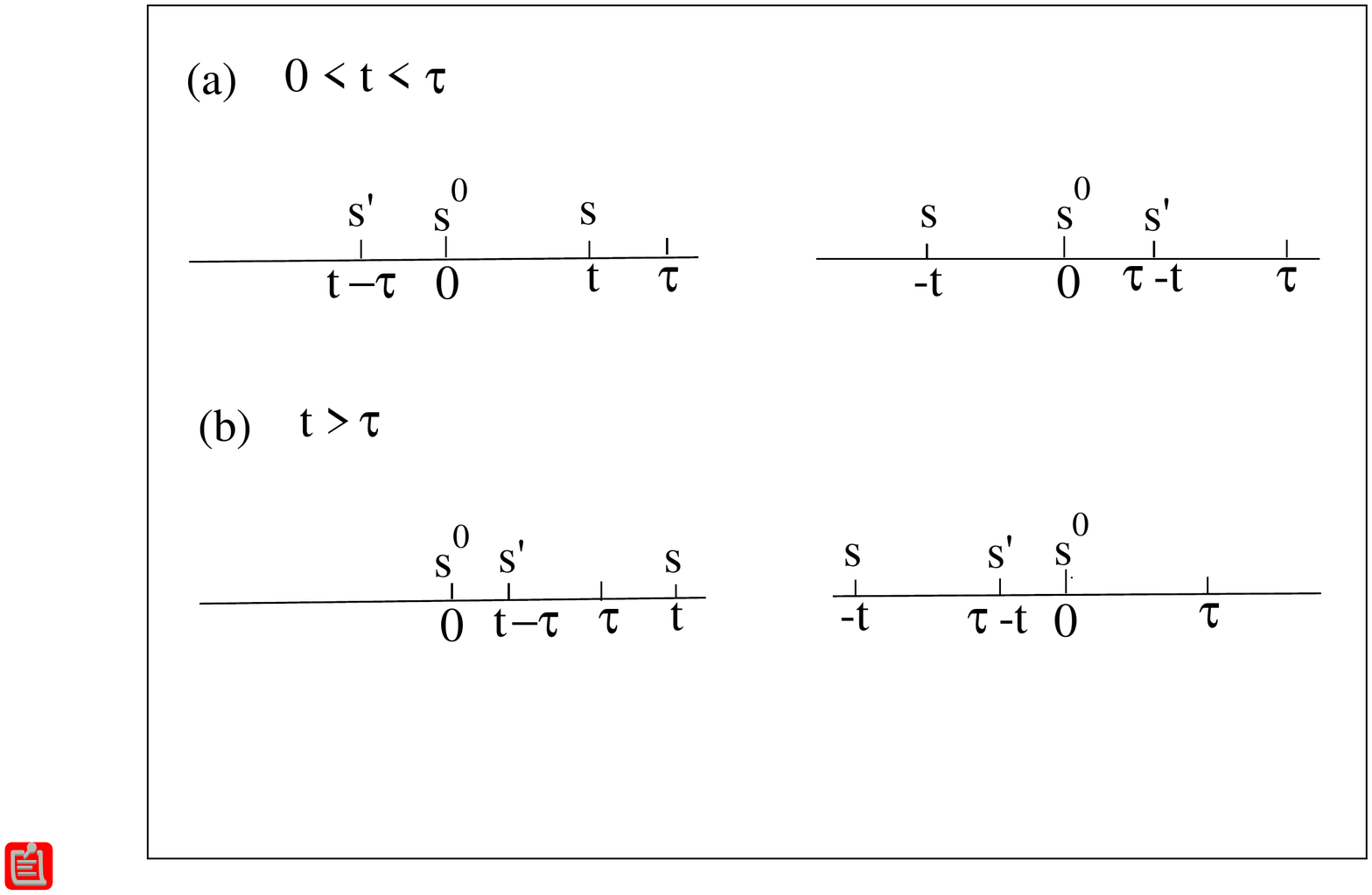}
 \caption{\label{Fig1} Sequence of states ${\bf s},{\bf s}'$ and ${\bf s}^{0}$ for (a) $0\le t\le \tau$  and  (b) $t > \tau$. On the right hand side, the direction of time is reversed.}
\end{center}
\end{figure}

We first compute the infinitesimal change $dp({\bf s}',t-\tau; {\bf s},t| {\bf s}^0)$ associated to an infinitesimal change $dt$.  Taking into account the fact that $t$ appears twice, we obtain 
\begin{align}
dp({\bf s}',t-\tau; {\bf s},t| {\bf s}^0)&=[p({\bf s}',t-\tau; {\bf s},t+dt\vert {\bf s}^0) -p({\bf s}',t-\tau; {\bf s},t \vert {\bf s}^0)]+[p({\bf s}',t+dt-\tau; {\bf s},t \vert {\bf s}^0) -p({\bf s}',t-\tau; {\bf s},t| {\bf s}^0)]\nonumber\\
&=[p({\bf s}',t-\tau; {\bf s},t+dt\vert {\bf s}^0) -p({\bf s}',t-\tau; {\bf s},t \vert {\bf s}^0)]+[p( {\bf s},-t;{\bf s}',\tau-t-dt\vert {\bf s}^0) -p({\bf s},-t;{\bf s}',\tau-t| {\bf s}^0)]\nonumber\\
&\equiv dp_1 +dp_2
\end{align}
where we have used Eq. (\ref{Timesym}) to reverse the direction of time in $dp_2$, the second term inside brackets (from now on we omit the subscript $\{\mbox{st}\}$ in the various probability distributions to simplify the notations). When $0\le t\le \tau$, the state ${\bf s}^0$ of the system at $t=0$  is irrelevant for calculating $dp_1/dt$ and $dp_2/dt$. Indeed, as illustrated in Fig. 1(a), one has $t-\tau<0<t$ in the first case and $-t<0<\tau -t$ in the second case. Since the switching rate at time $t$ only depends on the states at times $t$ and $t-\tau$  and the switching rate at time $\tau-t$ only depends on the states at times $\tau-t$ and $-t$, we simply have (cf. Eq. (\ref{Master1}) without the summations)
\begin{align}
\label{Eqdp}
\dot{p}({\bf s}',t-\tau; {\bf s},t| {\bf s}^0)&=\dot{p}_1+\dot{p}_2
\end{align}
with
\begin{align}
\label{Eqdp1}
\dot{p}_1=& T(-s_1\rightarrow s_1;{\bf s}')p({\bf s}',t-\tau;\{-s_1,s_2\},t| {\bf s}^0)
+ T(-s_2\rightarrow s_2;{\bf s}')p({\bf s}',t-\tau;\{s_1,-s_2\},t| {\bf s}^0) \nonumber\\
&-[T(s_1 \rightarrow -s_1;{\bf s}')+T(s_2\rightarrow -s_2;{\bf s}')]p({\bf s}',t-\tau;\{s_1,s_2\},t| {\bf s}^0) 
\end{align}
and 
\begin{align}
\label{Eqdp2}
\dot{p}_2=&-T(-s'_1\rightarrow s'_1;{\bf s})p({\bf s},-t;\{-s'_1,s'_2\},\tau -t| {\bf s}^0)
- T(-s'_2\rightarrow s'_2;{\bf s})p({\bf s},-t;\{s'_1,-s'_2\},\tau-t| {\bf s}^0) \nonumber\\
&+[T(s'_1 \rightarrow -s'_1;{\bf s})+T(s'_2\rightarrow -s'_2;{\bf s})]p({\bf s},-t;\{s'_1,s'_2\},\tau-t| {\bf s}^0) 
\end{align}
(note the change of sign in $\dot{p}_2$ in relation to the change of sign of $dt$).

On the other hand, when $t>\tau$, as illustrated in Fig. 1(b), one has $-t<\tau -t<0$ and a possible switching at time $\tau -t$ is conditioned by the fact that the system is in state ${\bf s}^0$ at $t=0$, which lies in the future. Therefore the simple probabilistic argument that leads to Eq. (\ref{Eqdp2}) is no more valid (for instance, it can be checked in the case $N=1$ that the corresponding equation does not yield the correct expression of the correlation function for $\tau<t\le 2\tau$ as computed in Ref.\cite{TP2001}). Despite our efforts, we have not succeeded so far to find the correct equation of motion of $p_{st}({\bf s}',t-\tau;{\bf s},t|{\bf s}^0)$ for $t>\tau$. 

To proceed and solve the set of coupled linear differential equations represented by Eqs. (\ref{Eqdp}-\ref{Eqdp2}), it is now necessary to make explicit the dependence on the $4$ configurations $\{1,1\}$, $\{1,-1\}$, $\{-1,1\}$, $\{-1,-1\}$  that we shall denote $A,B,C,D$, respectively. First, we note that all quantities must be invariant under particle exchange $s_1 \rightleftharpoons  s_2$ and sign exchange $+1\rightleftharpoons  -1$ (this symmetry is unbroken because no phase transition is expected). This readily implies that $p(B)=p(C)$ and  $p(A)=p(D)$ whence
\begin{align}
p(A)+p(B)=\frac{1}{2} \ .
\end{align}
Moreover, when calculating the conditional probabilities, the state ${\bf s}^{0}$ at $t=0$ may be chosen to be either A or B. Hence there are only $6$ distinct  functions $p({\bf s},t\vert{\bf s}^0)$, namely $p(A,t\vert A),p(B,t\vert A),p(D,t\vert A),p(A,t\vert B),p(B,t\vert B),p(C,t\vert B)$ and only $3$ of them are linearly independent. Indeed, from the equations expressing the conservation of probabilities 
\begin{align}
\sum_{{\bf s}} p({\bf s},t|{\bf s}^0)&=1\nonumber\\
\sum_{{\bf s}^0} p({\bf s},t|{\bf s}^0)p({\bf s}^0)&=p({\bf s}) \ ,
\end{align}
one can derive the $3$ relations
\begin{align}
\label{Eqprobflow}
&p(A)p(B,t\vert A)=p(B)p(A,t\vert B)\nonumber\\
&2p(B,t\vert A)+p(A,t\vert A)+p(D,t\vert A)=1\nonumber\\
&2p(A,t\vert B)+p(B,t\vert B)+p(C,t\vert B)=1 \ .
\end{align}
There are also only $2\times 10$ distinct functions among the $64$ conditional probabilities $p_{st}({\bf s}',t-\tau;{\bf s},t|{\bf s}^0)$ (which moreover are not  all linearly independent because of conservation of probabilities). Inserting the expression (\ref{Rates}) of the rates in Eqs. (\ref{Eqdp}-\ref{Eqdp2}), we thus obtain two sets of $10$ coupled linear differential equations which may be written as
\begin{align}
\label{EqdiffAB}
\dot{{\bf p}}_A(t)&={\bf \Gamma}_A {\bf p}_A(t)\nonumber\\
\dot{{\bf p}}_B(t)&={\bf \Gamma}_B {\bf p}_B(t)
\end{align}
with 
\begin{equation}
\label{Eqp1}
{\bf p}_A(t)=\begin{pmatrix}
p(B,t-\tau; B,t| A) \\
p(A,t-\tau; B,t| A) \\
p(D,t-\tau; B,t| A) \\
p(C,t-\tau; B,t| A) \\
p(B,t-\tau; A,t| A) \\
p(A,t-\tau; A,t| A) \\
p(D,t-\tau; A,t| A) \\
p(B,t-\tau; D,t| A) \\
p(A,t-\tau; D,t| A) \\
p(D,t-\tau; D,t| A) \\
\end{pmatrix} , \ {\bf p}_B(t)=\begin{pmatrix}
p(A,t-\tau; A,t| B) \\
p(B,t-\tau; A,t| B) \\
p(C,t-\tau; A,t| B) \\
p(D,t-\tau; A,t| B) \\
p(A,t-\tau; B,t| B) \\
p(B,t-\tau; B,t| B) \\
p(C,t-\tau; B,t| B) \\
p(A,t-\tau; C,t| B) \\
p(B,t-\tau; C,t| B) \\
p(C,t-\tau; C,t| B) \\ 
\end{pmatrix} \ ,
\end{equation}
 and ${\bf \Gamma}_A$, ${\bf \Gamma}_B$ are $10\times 10$ matrices whose expressions are given in Appendix B. The solution of these equations is 
\begin{align}
\label{Eqdiffp}
{\bf p}_A(t)&=e^{{\bf \Gamma}_A t} {\bf p}_A(0)\nonumber\\
{\bf p}_B(t)&=e^{{\bf \Gamma}_B t} {\bf p}_B(0) \ ,
\end{align}
and the problem reduces to the calculation of the eigenvalues and eigenvectors of ${\bf \Gamma}_A$ and ${\bf \Gamma}_B$, as detailed in Appendix B (the two matrices  have the same spectrum when Eq. (\ref{Eqgamma}) is satisfied).  It only remains to determine the initial conditions at $t=0$. One can see from Eqs. (\ref{Eqp1})  that  only $3$ components of the vectors  ${\bf p}_A(0)$ and  ${\bf p}_B(0)$ are nonzero, 
\begin{align}
\label{Eqp2}
p(A,-\tau; A,0| A)&=p(A,0; A,\tau| A)= p(A,\tau| A) \nonumber\\
p(B,-\tau; A,0| A)&=p(A,0; B,\tau| A)= p(B,\tau| A) \nonumber\\
p(D,-\tau; A,0| A)&=p(A,0; D,\tau| A)= p(D,\tau| A)
\end{align}
and 
\begin{align}
\label{Eqp3}
p(A,-\tau; B,0| B)&=p(B,0; A,\tau | B)= p(A,\tau| B) \nonumber\\
p(B,-\tau; B,0| B)&=p(B,0; B,\tau | B)= p(B,\tau| B) \nonumber\\
p(C,-\tau; B,0| B)&=p(B,0; C,\tau | B)= p(C,\tau| B) 
\end{align}
where we have used the time-reversal symmetry, Eq. (\ref{Timesym}). The $6$ quantities $p(A,\tau| A),p(B,\tau| A),...,p(C,\tau| B)$ are still unknown but they can be obtained by observing that Eqs.(\ref{Eqp2}-\ref{Eqp3}) also correspond to the values of  $p({\bf s}',t-\tau;{\bf s},t|{\bf s}^0)$ at $t=\tau$. Therefore, they are also given by
 \begin{align}
\label{Eqselfcons}
{\bf p}_A(\tau)&=e^{{\bf \Gamma}_A \tau} {\bf p}_A(0) \nonumber\\
{\bf p}_B(\tau)&=e^{{\bf \Gamma}_B \tau} {\bf p}_B(0) \ ,
\end{align}
which yields two sets of self-consistency equations  whose solution is given in Appendix B. Interestingly, these equations have nontrivial solutions only when Eq. (\ref{Eqgamma}) is satisfied. Therefore this relation again appears as a necessary condition for the problem under study to be integrable.

Knowing these quantities we can calculate all the components of ${\bf p}_A(t)$ and  ${\bf p}_B(t)$ (these functions  are sums of six exponential factors -see Appendix B- and the explicit expressions are not given here for the sake of brevity), then the probabilities $p({\bf s},t| {\bf s}^0)=\sum_{{\bf s}'}p({\bf s}',t-\tau;{\bf s},t|{\bf s}^0)$, and finally $p(A)$ and the correlation functions $\psi_s(t), \psi_c(t)$.  We have verified that the probabilities $p({\bf s},t| {\bf s}^0)$ obtained in this way satisfy  the equations of motion (\ref{Master11}), which shows that the whole calculation is indeed consistent. 

The correlation functions are calculated using
\begin{align}
\psi_s(t)&=2p(A)[p(A,t\vert A)-p(D,t\vert A)]+2p(B)[p(B,t\vert B)-p(C,t\vert B)]\nonumber\\
\psi_c(t)&=2p(A)[p(A,t\vert A)-p(D,t\vert A)]-2p(B)[p(B,t\vert B)-p(C,t\vert B)] \ .
\end{align}

\begin{figure}[hbt]
\begin{center}
\includegraphics[width=10cm]{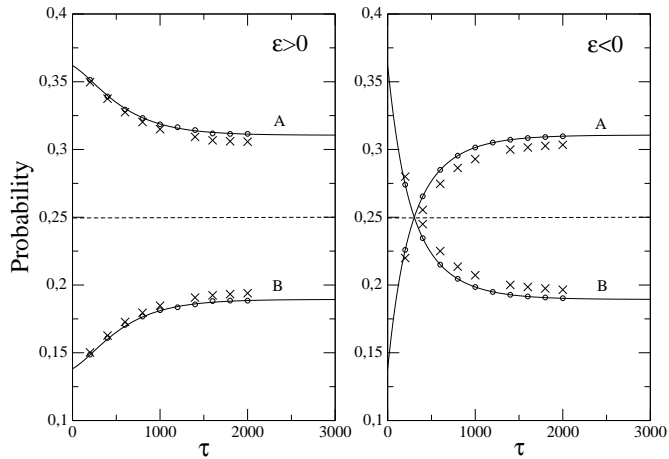}
 \caption{\label{Fig2} Stationary occupation probabilities $p(A)$ and $p(B)=1/2-p(A)$ as a function of the time delay $\tau$ for $D=0.05$, $\epsilon=0.05$ (left panel) and $\epsilon=-0.05$ (right panel). Theoretical results (solid lines) are compared to numerical simulations of the stochastic process  described by the rates (\ref{RatesN2}) (circles) and simulations of the original Langevin equations (\ref{Eq1}) (crosses).}
\end{center}
\end{figure}
To test the validity of our analytical results, we compared them to numerical simulations of the stochastic two-state process with switching rates obtained from the Kramers relations (\ref{RatesN2}) using  $D=0.05$ and  $\epsilon=\pm 0.05$. For $\epsilon>0$ this yields $\gamma_0=0.000578,\gamma_2=0.003965$,  and $\gamma_1=0.001516$ (for $\epsilon<0$,  the values of $\gamma_0$ and $\gamma_2$ are interchanged). $\gamma_1$ was actually adjusted to the value $\sqrt{\gamma_0\gamma_2}=0.001514$ so to exactly satisfy Eq. (\ref{Eqgamma}).
The numerical simulations were carried out using a time-step $\Delta t=0.01$ and averages were taken over a run of $16\times 10^9$ steps, discarding the first $10^9$ steps. We also performed simulations of the Langevin dynamics described by Eq. (\ref{Eq1}) using Euler method. 

Fig. \ref{Fig2} shows the dependence of the stationary probabilities $p(A)$ and $p(B)=1/2 -p(A)$ on the time delay $\tau$. As can be seen, the theory is in perfect agreement with the simulations of the stochastic process. The agreement with the Langevin dynamics is also reasonably good for $\epsilon>0$ whereas some systematic deviations are observed for $\epsilon<0$. There are indeed more fluctuations in the latter case and the random variables $x_1(t)$ and $x_2(t)$ often take intermediate values between $-1$ and $+1$ making the two-state model less accurate.

\begin{figure}[hbt]
\begin{center}
\includegraphics[width=12cm]{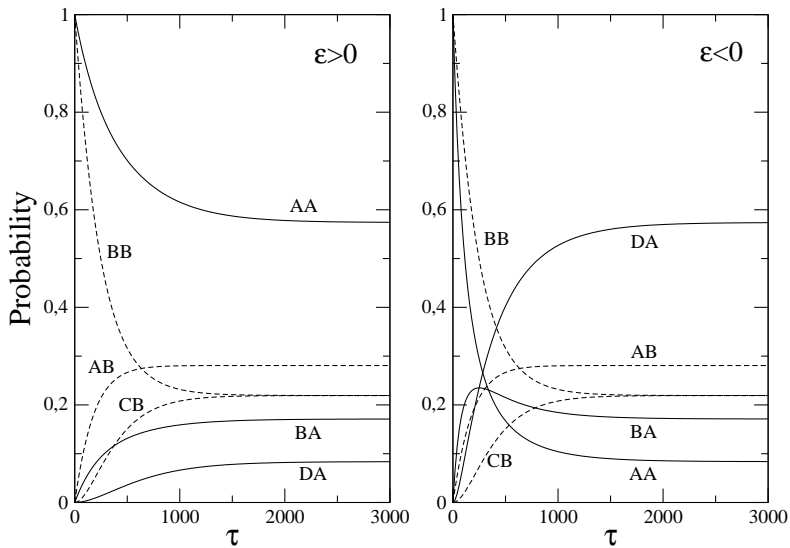}
 \caption{\label{Fig3} Probabilities $p(A,\tau| A),p(B,\tau| A),p(D,\tau|A),p(A,\tau|B),p(B,\tau|B),p(C,\tau|B)$ as a function of $\tau$ for $\epsilon=0.05$ (left panel) and $\epsilon=-0.05$ (right panel).}
\end{center}
\end{figure}

It is worth noting that $p(A)$ goes to a nontrivial limit when $\tau \rightarrow \infty$ (given analytically by Eq. (\ref{tauinf})). Indeed, the trivial value $1/4$ would be obtained by naively assuming that the events at $t$ and $t-\tau$  in the master equations (\ref{Master1}) can be decoupled when $\tau$ is much larger than the other characteristic times of the system. The actual limit  is larger than $1/4$  and is the same for $\epsilon>0$ and $\epsilon<0$, implying that the two configurations A and D where the two spins are in the same state are favored {\it whatever} the sign of the feedback (this contrasts with the behavior for small time delays where the ratio $p(A)/p(B)$ is larger or smaller than $1$ depending on the sign of $\epsilon$: this result can be easily understood by referring to the case $\tau =0$ where the process becomes Markovian). To understand the behavior for larger time delays, it is instructive to consider  the conditional probabilities  $p(A,\tau| 
 A),p(B,\tau| A),...,p(C,\tau|B)$ whose dependence on $\tau$ is shown in Fig. \ref{Fig3}.

We note in particular that $p(A,\tau| A)$ is larger or smaller than $p(D,\tau| A)$ depending on the sign of $\epsilon$. This is not surprising as the global coupling $\epsilon X(t-\tau)$ increases or decreases the probability for an element to be at time $t$ in the same potential well in which the majority of elements were at time $t-\tau$ depending on whether $\epsilon$ is positive or negative\cite{HT2003}. Since configurations B and C do not contribute to the mean field $X$, the force that determines the evolution of the state $A$ at time $t$ is (on average) $\epsilon <X(t-\tau)>=\epsilon [p(A,t-\tau| A,t)- p(D,t-\tau| A,t)]$ which is equal to  $\epsilon [p(A,\tau| A)- p(D,\tau| A)]$ in the stationary state. This force is thus positive on average whatever the sign of $\epsilon$ and the configuration $A$ is stabilized. On the other hand, for configuration $B$, the average force is equal to  $\epsilon [p(A,\tau| B)- p(D,\tau| B)]$ which is zero by symmetry. Although this is a
  mean-field argument, it qualitatively explains why configurations A or D are more probable than B or C for both positive and negative feedbacks, as observed in Fig. \ref{Fig2} for large $\tau$.

The behavior of the occupation probabilities as a function of the time delay observed in Figs. {\ref{Fig2} and {\ref{Fig3} (note also the non-monotonic variation of $p(B,\tau| A)$ in this latter figure) suggests that the dependence on the coupling strength at fixed $\tau$ may be also interesting. This is illustrated in Fig. \ref{Fig4} where the ratio $p(B)/p(A)$ computed from Eq. (\ref{Finalresult}) is plotted as a function of $\epsilon$ for $\tau=500$. We see that there is a range of negative values of $\epsilon$ where $p(B)$ is slightly larger than $p(A)$ with a maximum occurring at $\epsilon  \approx -0.012$.

\begin{figure}[hbt]
\begin{center}
\includegraphics[width=9cm]{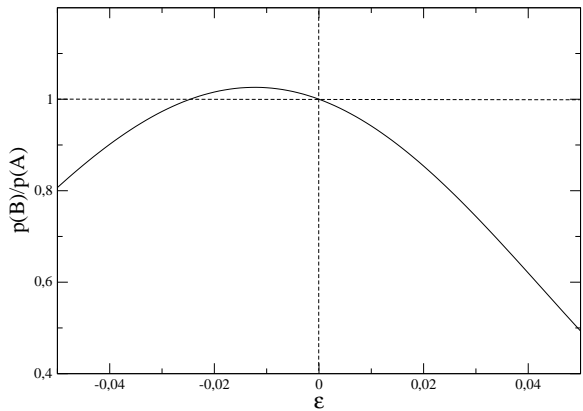}
 \caption{\label{Fig4} The ratio of the stationary occupation probabilities $p(B)/p(A)$ as a function of the coupling strength $\epsilon$ for $D=0.05$ and $\tau=500$. This quantity is computed from the analytical expression of $p(A)$ given in Appendix B.}
\end{center}
\end{figure}

\begin{figure}[hbt]
\begin{center}
\includegraphics[width=11cm]{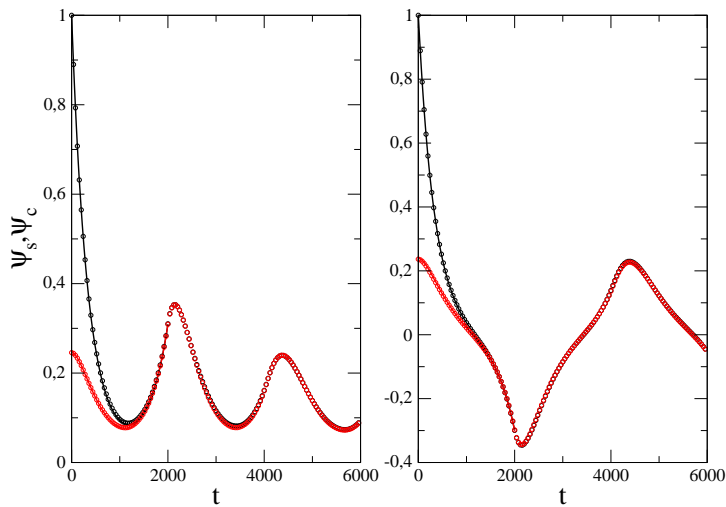}
 \caption{\label{Fig5} (Color on line) Self (black) and cross (red) time correlation functions $\psi_s(t)$ and $\psi_c(t)$ for $D=0.05$, $\tau=2000$, $\epsilon=0.05$ (left panel) and $\epsilon=-0.05$ (right panel). For $0\le t\le \tau$, the theoretical results (solid lines) are compared to numerical simulations of the stochastic process (circles).}
\end{center}
\end{figure}
Finally, the self and cross time correlation functions for $\tau =2000$ are shown in  Fig. \ref{Fig5}. For $\vert \epsilon\vert=0.05$, this value of $\tau$ is slightly larger than the largest characteristic time of the system ($t_0=1/\gamma_0\approx 1729$ for $\epsilon>0)$. There is again perfect agreement between theory and simulations of the stochastic two-state process in the interval $[0,\tau]$ where the analytical results are available (note that $\psi_c(0)=4p(A)-1$). For $\epsilon>0$ both functions are positive with maxima at $t\approx n\tau$ whereas for $\epsilon<0$ the peaks at $t\approx n\tau$ have alternating signs (of course, the functions go to $0$ as $t\rightarrow \infty$). Moreover, the peaks are always delayed with respect to $n\tau$, a behavior that was already observed in the case $N=1$\cite{TP2001}.

\section{ Summary and conclusion}

We have studied a system of two bistable elements with a global time-delayed coupling using a two-state model where the dynamics is described by delay-differential master equations. In general, due to the non-Markovian nature of the dynamics, this set of equations is not closed since one-time occupation probabilities depend on two-time probabilities, two-time probabilities on three-time probabilities, etc... We have shown, however, that one can close this infinite hierarchy and derive analytical expressions for the occupation probabilities and the correlation functions in the stationary state provided the stochastic process is time-symmetric. This  property only holds when some relationship between the different switching rates is satisfied, which is approximately the case when the rates are described by Kramers's theory in the limit of very small coupling.  It is rather obvious that the explicit demonstration presented in this  work can be generalized to a larger number of interacting elements, which opens the way for a full (though admittedly intricate) analytical description. We stress, however, that our solution for $N=2$ is still incomplete since the analytical expressions of the time correlation functions are only valid for $t$ smaller than the time delay $\tau$, which precludes the calculation of the power spectrum as was done for $N=1$\cite{TP2001} or $N\rightarrow\infty$\cite{HT2003}. Obtaining an analytical description valid for all times is clearly the most challenging task for future work. It would also be interesting to compute the distribution of residence times along the lines of Ref.\cite{M2003}. Our results for $N=2$ show that the main effect of the global coupling when the time delay is not too small is to increase the probability for the elements to be in the same potential well, whatever the sign of the feedback coupling. As $N$ increases, one should start to observe on a certain time-scale the nontrivial behavior exhibited in the thermodynamic limit\cite{HT2003}. For instance, since there exists an ordered phase with a non-zero stationary mean field $X$ for a sufficiently large positive coupling, one should see $X$  switching between these non-zero values with a rate decreasing with $N$ \cite{PZC2002}.

\appendix 

\section{Time-symmetry in the stationary state}

In this appendix we show that the stochastic process described by the hopping rates (\ref{RatesN2}) is statistically time-symmetric in the stationary state when relation (\ref{Eqgamma}) is satisfied.
Time reversibility is not at all obvious because stochastic processes with delay are non-Markovian by definition. However,  the probability of observing a given path during a time interval $[t,t+\tau]$ only depends on the realization of the process during the preceding time interval $[t-\tau,t]$ , a property that we shall use repeatedly in the following (more generally, this allows one to describe delay processes in terms of Markov processes at the price of enlarging the space of random variables, see {\it e.g.} \cite{F2002,F2005}). 

For this purpose we consider a discrete time approximation of the original process by defining a set of equidistant observation times $n\Delta t$ with a time-step
\begin{eqnarray}
\Delta t=\frac{\tau}{M} 
\end{eqnarray}    
where $M$ is some large integer. The discrete time process converges to the original continuous one as $M\to \infty$.
The random path $({\bf s}_{t_0}={\bf s}^{(0)},{\bf s}_{t_1}={\bf s}^{(1)},...,{\bf s}_{t_M}={\bf s}^{(M)})$ at times $t_0,t_1=t_0+\Delta t,...,t_{M}=t_0+\tau$ is then denoted $S^{0,M}$ and  the conditional probability of observing the  path $S^{M+1,2M+1}$ given the path $S^{0,M}$  is denoted by $P(S^{M+1,2M+1}|S^{0,M})$ (as we only study the stationary state, we can set $t_0=0$ without loss of generality). We shall first prove that
\begin{eqnarray}
\label{EqTimerev}
p_{st}(S^{0,M})=p_{st}(S^{M,0}) \ ,
\end{eqnarray}    
and then extend the proof  to a path of arbitrary duration. To simplify the notation, the subscript $st$ is dropped in the following and the time reversal of a path $S$ (for instance $S^{M,0}$) is denoted $\bar{S}$.

\subsection{The case $N=1$}

\begin{figure}[hbt]
\begin{center}
\includegraphics[width=7cm]{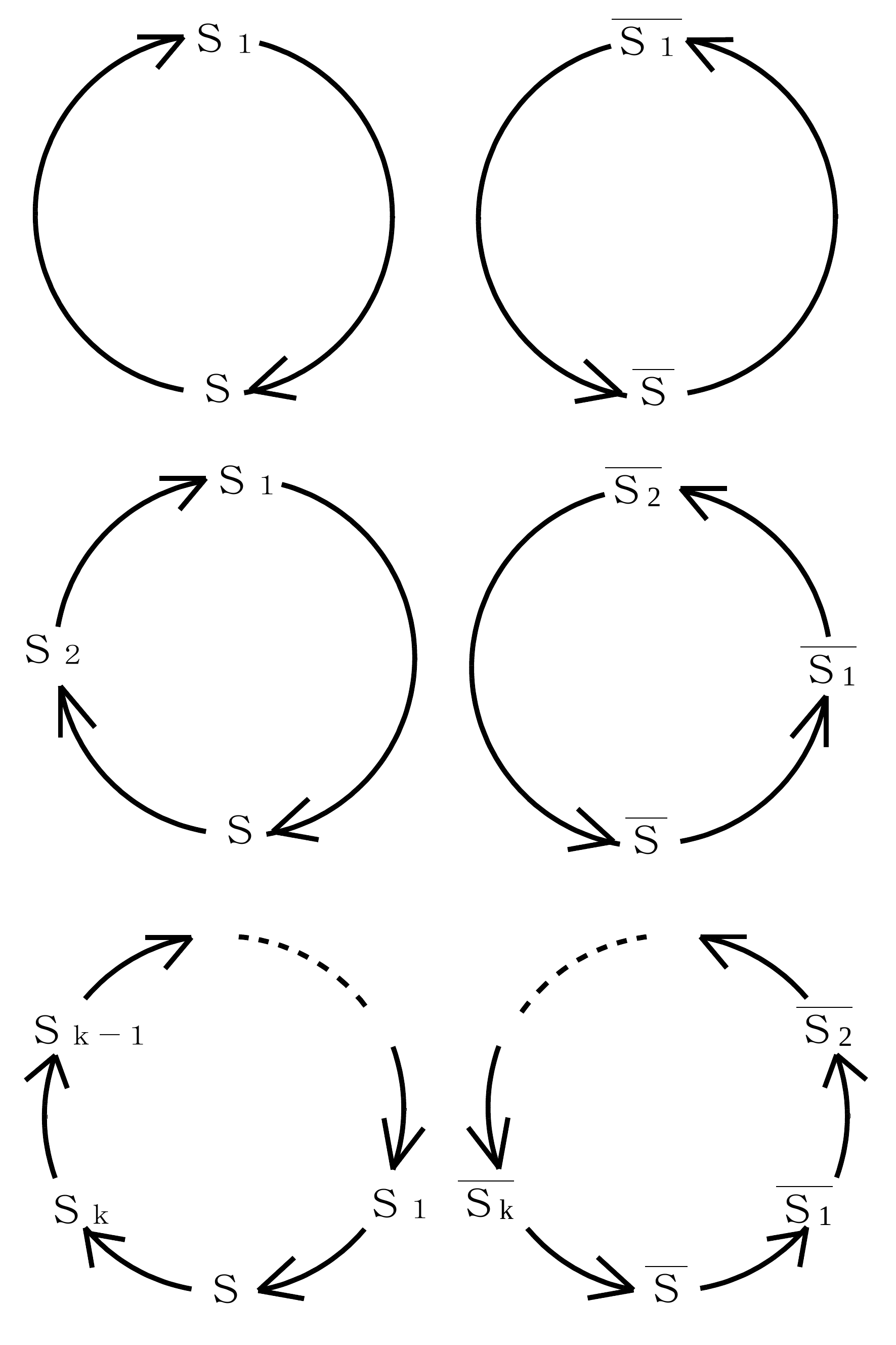}
 \caption{\label{Process} Description of the sequences of trajectories of increasing length that are used in the proof of Eq. (\ref{EqTimerev}). In each case, the first and last trajectories of duration $\tau$ are identical. On the right-hand side, the time-reversal of the sequences is considered.}
\end{center}
\end{figure}
As a warm-up, we first consider the case $N=1$. As we shall see, time symmetry is always valid,  even without taking the continuous limit $\Delta t \to 0$. This contrasts with the case $N=2$ that is studied in the next subsection. 

To prove Eq. (\ref{EqTimerev}) we consider sample paths of increasing length such that the path in the last time interval of duration $\tau$ is the same as in the first interval $[0,\tau]$, as illustrated schematically in Fig. \ref{Process}. We thus first start with a sample path $S\rightarrow S_1\rightarrow S$ of duration $3\tau+2\Delta t$ that reads $S^{0,3M+2}=(s_{t_0}=s^{(0)},s_{t_1}=s^{(1)},...,s_{t_M}=s^{(M)},s_{t_{M+1}}=s^{(M+1)},s_{t_{M+2}}=s^{(M+2)},...,s_{t_{2M+1}}=s^{(2M+1)},s_{t_{2M+2}}=s^{(0)},s_{t_{2M+3}}=s^{(1)},...,s_{t_{3M+2}}=s^{(M)})$ (here $S$  (resp. $S_1$) is a short-hand notation for the sample path in the intervals $[0,\tau]$ and $[2\tau+2\Delta t,3\tau+2\Delta t]$ (resp. $[\tau+\Delta t,2\tau+\Delta t]$)). We then consider the time reverse of this path,  {\it i.e.} $\overline{S} \rightarrow \overline{S}_1\rightarrow \overline{S}$ and prove that
\begin{eqnarray}
\label{Eqstep1}
P(S_1|S)P(S|S_1)=P(\overline{S_1}|\overline{S})P(\overline{S}|\overline{S_1}) \ .
\end{eqnarray}
The crucial point is that the probability of observing $s^{(i)}$ only depends on $s^{(i-1)}$, the value of the spin at the preceding time step, and $s^{(i-M-1)}$, the value of the spin $M+1$ steps earlier. Hence 
\begin{equation}
\label{process1}
P(S_1|S)=\prod_{i=M+1}^{2M+1}p(s^{(i)}|s^{(i-1)},s^{(i-M-1)}) \ ,
\end{equation} 
where the probabilities $p(s^{(i)}|s^{(i-1)},s^{(i-M-1)})$ are given by the transition rates defined by Eq. (\ref{RatesN1})
\begin{align}
 p(+1|+1,+1)&=p(-1|-1,-1)=1-\gamma_1\Delta t \nonumber\\ 
 p(-1|+1,+1)&=p(+1|-1,-1)=\gamma_1\Delta t \nonumber\\ 
 p(+1|+1,-1)&=p(-1|-1,+1)=1-\gamma_2\Delta t \nonumber\\ 
 p(-1|+1,-1)&=p(+1|-1,+1)=\gamma_2\Delta t  \ .
\end{align}
Eq. (\ref{process1}) can thus be recast as 
\begin{eqnarray}
P(S_1|S)= (1-p_1)^{m_1}p_1^{n_1}(1-p_2)^{m_2}p_2^{n_2} \ ,
\end{eqnarray}
where $p_1=\gamma_1 \Delta t$, $p_2=\gamma_2 \Delta t$, and $m_1,m_2,n_1,n_2$ count how many times the factors $1-p_1,1-p_2,p_1,p_2$ appear in the right-hand side of Eq. (\ref{process1}). These numbers are readily obtained by introducing the binary variables 
\begin{align}
a_{\nu}=\frac{s^{(\nu)}+1}{2},\ b_{\nu}=\frac{s^{(\nu+M+1)}+1}{2},\ \ \  0\leq \nu\leq M
\end{align}
which yields 
\begin{align}
m_1&=\sum_{\nu=0}^{M} [a_{\nu}b_{\nu-1}b_{\nu}+(1-a_{\nu})(1-b_{\nu-1})(1-b_{\nu})] \nonumber\\
n_1&=\sum_{\nu=0}^{M} [a_{\nu}b_{\nu-1}(1-b_{\nu})+(1-a_{\nu})(1-b_{\nu-1})b_{\nu}]\nonumber\\
m_2&=\sum_{\nu=0}^{M} [a_{\nu}(1-b_{\nu-1})(1-b_{\nu})+(1-a_{\nu})b_{\nu-1}b_{\nu}] \nonumber\\
n_2&=\sum_{\nu=0}^{M} [a_{\nu}(1-b_{\nu-1})b_{\nu}+(1-a_{\nu})b_{\nu-1}(1-b_{\nu})] \ . 
\end{align}
with the convention $b_{-1}\equiv a_M$.
Similarly, we can also write the left and right-hand sides of Eq. (\ref{Eqstep1}) as 
$(1-p_1)^{k_1}p_1^{l_1}(1-p_2)^{k_2}p_2^{l_2}$ and $(1-p_1)^{\overline{k}_1}p_1^{\overline{l}_1}(1-p_2)^{\overline{k}_2}p_2^{\overline{l}_2}$ respectively. The explicit calculation readily shows  that $k_1=\overline{k}_1, k_2=\overline{k}_2,l_1=\overline{l}_1$ and $l_2=\overline{l}_2$, which proves Eq. (\ref{Eqstep1}).

We then intercalate between $S$ and $S_1$ an additional path $S_2$ of duration $\tau$, as shown in Fig \ref{Process}, and we repeat the same calculation. After some algebra we find that
\begin{equation}
P(S_2|S)P(S_1|S_2)P(S|S_1)=P(\overline{S_1}|\overline{S})P(\overline{S_2}|\overline{S_{1}}) 
P(\overline{S}|\overline{S_2}) \ .
\end{equation}
It is then straightforward to prove  that 
\begin{eqnarray}
\label{Eqpk}
P(S_{k}|S)P(S_{k-1}|S_{k})\times \dots \times P(S|S_1)=P(\overline{S_1}|\overline{S})P(\overline{S_2}|\overline{S_1}) 
\times \dots \times P(\overline{S}|\overline{S_{k}}) 
\end{eqnarray}
for an arbitrary integer $k\ge 1$. Summing over the intermediate states $(S_1,S_2,...,S_k)$ we then find that \begin{eqnarray}
P(S, k|S)=P(\overline S,k|\overline S) \ ,
\end{eqnarray}
where $P(S, k|S)$ is the conditional probability to realize the path $S$ after $k$ intermediate paths of duration $\tau$, given the path $S$ in the first time interval $[0,\tau]$.  Assuming ergodicity, we can forget the initial condition in the limit $k\to \infty$, which yields Eq. (\ref{EqTimerev}).

The proof can easily be extended to a path of duration $n\tau$ (and more generally of any duration). Take for instance $n=3$ and consider the path $S_c\rightarrow S_k\rightarrow S_{k-1}...\rightarrow S_1\rightarrow S_a\rightarrow S_b\rightarrow S_c$ where each individual path is of duration $\tau$. Then from Eq. (\ref{Eqpk}) we have
\begin{align}
P(S_{k}|S_c)P(S_{k-1}|S_{k})\times \dots \times P(S_a|S_1)P(S_b\vert S_a)P(S_c\vert S_b)&=P(\overline{S_b}|\overline{S_c})P(\overline{S_a}|\overline{S_b}) P(\overline{S_1}|\overline{S_a}) \nonumber\\
&\times \dots \times P(\overline{S_k}|\overline{S_{k-1}}) P(\overline{S_c}|\overline{S_{k}}) \ .
\end{align}
Then summing over all intermediate states $(S_1,S_2...S_k)$ we obtain
\begin{align}
P(S_a, k(\tau+\Delta t)|S_c)P(S_b\vert S_a)P(S_c\vert S_b)=P(\overline{S_b}|\overline{S_c})P(\overline{S_a}|\overline{S_b}) P(\overline{S_c},k(\tau+\Delta t)|\overline{S_a})  
\end{align}
where $P(S_a, k(\tau+\Delta t)|S_c)$ (resp. $P(\overline{S_c},k(\tau+\Delta t)|\overline{S_a})$ ) is the conditional probability to realize the path $S_a$ (resp. $\overline{S_c}$) after $k$ intervals of duration $\tau+\Delta t$ given the path $S_c$ (resp. $\overline{S_a}$). In the limit $k \to \infty$, $P(S_a, k(\tau+\Delta t)|S_c) \to p(S_a)$ and $P(\overline{S_c},k(\tau+\Delta t)|\overline{S_a})\to p(\overline{S_c})$, whence
\begin{align}
p(S_a)P(S_b\vert S_a)P(S_c\vert S_b)=P(\overline{S_b}|\overline{S_c})P(\overline{S_a}|\overline{S_b}) p(\overline{S_c})  
\end{align}
which may be recast as
\begin{align}
p(S_a)P(S_b,S_c\vert S_a)=P(\overline{S_a},\overline{S_b}|\overline{S_c})p(\overline{S_c})
\end{align}
and finally
\begin{align}
p(S_a,S_b,S_c)=p(\overline{S_c},\overline{S_b},\overline{S_a}) 
\end{align}
where $(S_a,S_b,S_c)$ is a path of duration $3\tau$.
\subsection{The case $N=2$}

The proof for the case $N=2$ is conducted along the same lines, starting with the sample  path $S\rightarrow S_1\rightarrow S$ of duration $3\tau+2\Delta t$, where ${\bf s}$ is now the two-component vector $(s_1,s_2)$ and the rates are given by  Eq. (\ref{Rates}). Eq. (\ref{process1}) thus becomes 
\begin{eqnarray}
\label{process2}
P(S_1\vert S)=\prod_{i=M+1}^{2M+1}p(s_1^{(i)}|s_1^{(i-1)},{\bf s}^{(i-M-1)})
p(s_2^{(i)}|s_2^{(i-1)},{\bf s}^{(i-M-1)}) \ ,
\end{eqnarray} 
with
\begin{align}
 p(+1|+1,A)&=p(-1|-1,D)=1-\gamma_0\Delta t \nonumber\\ 
p(-1|+1,A)&=p(+1|-1,D)=\gamma_0\Delta t\nonumber\\ 
 p(+1|-1,A)&=p(-1|+1,D)=\gamma_2\Delta t  \nonumber\\ 
 p(-1|-1,A)&=p(+1|+1,D)=1-\gamma_2\Delta t\nonumber\\ 
 p(-1|+1,B)&=p(+1|-1,B)=p(+1|-1,C)=p(-1|+1,C)=\gamma_1\Delta t\nonumber\\  
 p(+1|+1,B)&=p(-1|-1,B)=p(+1|+1,C)=p(-1|-1,C)=1-\gamma_1\Delta t
\end{align}
Introducing again the binary variables
\begin{align}
a_{1,2}^{\nu}&=\frac{s_{1,2}^{(\nu)}+1}{2}, \ b_{1,2}^{\nu}=\frac{s_{1,2}^{(\nu+M+1)}+1}{2},\ \ \  0\leq \nu\leq M \ ,
\end{align}
we then find
\begin{eqnarray}
P(S_1\vert S)= (1-p_0)^{m_0}p_0^{n_0}(1-p_1)^{m_1}p_1^{n_1}(1-p_2)^{m_2}p_2^{n_2}
\end{eqnarray} 
where $p_0=\gamma_0 \Delta t,p_1=\gamma_1 \Delta t,p_2=\gamma_2 \Delta t,$ and $m_0,m_1,m_2$,  $n_0,n_1,n_2$ are given by  
\begin{align}
m_0&=\sum_{\nu=0}^{M} \big\{ a^{\nu}_1a^{\nu}_2(b^{\nu-1}_1b^{\nu}_1+b^{\nu-1}_2b^{\nu}_2) +(1-a^{\nu}_1)(1-a^{\nu}_2)[(1-b^{\nu-1}_1)(1-b^{\nu}_1) +(1-b^{\nu-1}_2)(1-b^{\nu}_2)] \big\}\nonumber\\
n_0&=\sum_{\nu=0}^{M}  \big\{ a^{\nu}_1a^{\nu}_2[b^{\nu-1}_1(1-b^{\nu}_1)+b^{\nu-1}_2(1-b^{\nu}_2)] 
+(1-a^{\nu}_1)(1-a^{\nu}_2)[(1-b^{\nu-1}_1)b^{\nu}_1+(1-b^{\nu-1}_2)b^{\nu}_2]  \big\}\nonumber\\
m_1&=\sum_{\nu=0}^{M} \big \{ [a^{\nu}_1(1-a^{\nu}_2)+(1-a^{\nu}_1)a^{\nu}_2] [b^{\nu-1}_1b^{\nu}_1+b^{\nu-1}_2b^{\nu}_2+(1-b^{\nu-1}_1)(1-b^{\nu}_1)+(1-b^{\nu-1}_2)(1-b^{\nu}_2)]  \big\}\nonumber\\
n_1&=\sum_{\nu=0}^{M} \big \{ [a^{\nu}_1(1-a^{\nu}_2)+(1-a^{\nu}_1)a^{\nu}_2] [b^{\nu-1}_1(1-b^{\nu}_1)+b^{\nu-1}_2(1-b^{\nu}_2) +(1-b^{\nu-1}_1)b^{\nu}_1+(1-b^{\nu-1}_2)b^{\nu}_2] \big \}\nonumber\\
m_2&=\sum_{\nu=0}^{M}  \big\{ (1-a^{\nu}_1)(1-a^{\nu}_2)(b^{\nu-1}_1b^{\nu}_1+b^{\nu-1}_2b^{\nu}_2) 
+a^{\nu}_1a^{\nu}_2 ([(1-b^{\nu-1}_1)(1-b^{\nu}_1)+(1-b^{\nu-1}_2)(1-b^{\nu}_2)]  \big\}\nonumber\\
n_2&=\sum_{\nu=0}^{M}  \big\{ (1-a^{\nu}_1)(1-a^{\nu}_2) [b^{\nu-1}_1(1-b^{\nu}_1)+b^{\nu-1}_2(1-b^{\nu}_2)] +a^{\nu}_1a^{\nu}_2[(1-b^{\nu-1}_1)b^{\nu}_1+(1-b^{\nu-1}_2)b^{\nu}_2] \big \} \ .
\end{align}
with $b^{-1}_1\equiv a^M_1$ and $b^{-1}_2\equiv a^M_2$ (note that $\nu$ is here an index and not an exponent).
The first task is to prove Eq. (\ref{Eqstep1}) as in the case $N=1$.  This amounts to proving that the ratio
\begin{eqnarray}
R\equiv \frac{P(S_1|S)P(S|S_1)}{P(\overline{S_1}|\overline{S})P(\overline{S}|\overline{S_1})}
\end{eqnarray}
is equal to $1$. After lengthy calculations we find that
\begin{equation}
R=\prod_{j=0}^2 (1-p_j)^{k_j}p_j^{l_j} \, 
\end{equation}
with
\begin{align}
k_0&=k_2=-\frac{k_1}{2} \nonumber\\
l_0&=l_2=-\frac{l_1}{2} \ ,
\end{align}
\begin{align}
\label{Eqm1}
k_1&=\sum_{\nu=0}^{M} \big\{ [b^{\nu+1}_1-b^{\nu}_1+b^{\nu+1}_2-b^{\nu}_2 +2(b^{\nu+1}_1b^{\nu+1}_2-b^{\nu}_1b^{\nu}_2)] [a^{\nu}_1a^{\nu+1}_1+a^{\nu}_2a^{\nu+1}_2+(1-a^{\nu}_1)(1-a^{\nu+1}_1) +(1-a^{\nu}_2)(1-a^{\nu+1}_2)] \nonumber\\
&+[a^{\nu+1}_1-a^{\nu}_1+a^{\nu+1}_2-a^{\nu}_2 +2(a^{\nu+1}_1a^{\nu+1}_2-a^{\nu}_1a^{\nu}_2)] [b^{\nu}_1b^{\nu+1}_1+b^{\nu}_2b^{\nu+1}_2+(1-b^{\nu}_1)(1-b^{\nu+1}_1) +(1-b^{\nu}_2)(1-b^{\nu+1}_2)] \big\} \nonumber\\
l_1&=\sum_{\nu=0}^{M} \{ [b^{\nu+1}_1-b^{\nu}_1+b^{\nu+1}_2-b^{\nu}_2 +2(b^{\nu+1}_1b^{\nu+1}_2-b^{\nu}_1b^{\nu}_2)] [a^{\nu}_1(1-a^{\nu+1}_1)+a^{\nu}_2(1-a^{\nu+1}_2) 
+(1-a^{\nu}_1)a^{\nu+1}_1+(1-a^{\nu}_2)a^{\nu+1}_2]\nonumber\\
& +[a^{\nu+1}_1-a^{\nu}_1+a^{\nu+1}_2-a^{\nu}_2 +2(a^{\nu+1}_1a^{\nu+1}_2-a^{\nu}_1a^{\nu}_2)] [b^{\nu}_1(1-b^{\nu+1}_1)+b^{\nu}_2(1-b^{\nu+1}_2) +(1-b^{\nu}_1)b^{\nu+1}_1+(1-b^{\nu}_2)(1-b^{\nu+1}_2)] \} 
\end{align}
with the convention  $a^{M+1}_{1,2}\equiv b^0_{1,2}$ and $b^{M+1}_{1,2}\equiv a^0_{1,2}$. Hence
\begin{align}
R&=\big[\frac{(1-p_1)^2}{(1-p_0)(1-p_2)}\big]^{k_1/2}\big[\frac{p_1^2}{p_0p_2}\big]^{l_1/2} \ .
\end{align}
The two numbers $k_1$ and $l_1$ are not zero and therefore  $R$ is not equal to $1$ at this stage. However, one can easily convince oneself that $k_1$ and $l_1$ are related to the number of switchings of the system during the time interval $\tau$ (for instance, we see in the expression of $k_1$ that the first term inside brackets is zero if $b_1^{\nu+1}=b_1^{\nu}$ and 
$b_2^{\nu+1}=b_2^{\nu}$, which means that the state of the system has not changed in the corresponding time step $\Delta t$). Since this number of switchings  remains finite in the continuous limit $\Delta t\to 0$, we then have  
\begin{align}
R&=\big[\frac{(1-\gamma_1\Delta t)^2}{(1-\gamma_0\Delta t)(1-\gamma_2\Delta t)}\big]^{k_1/2} \big[\frac{\gamma_1^2}{\gamma_0\gamma_2}\big]^{l_1/2} \rightarrow (\frac{\gamma_1^2}{\gamma_0\gamma_2})^{l_1/2} 
\end{align}
when $\Delta t\to 0$, and we conclude that  the transition rates must satisfy Eq. (\ref{Eqgamma}) in order to have $R=1$.

\begin{figure}[hbt]
\begin{center}
\includegraphics[width=9cm]{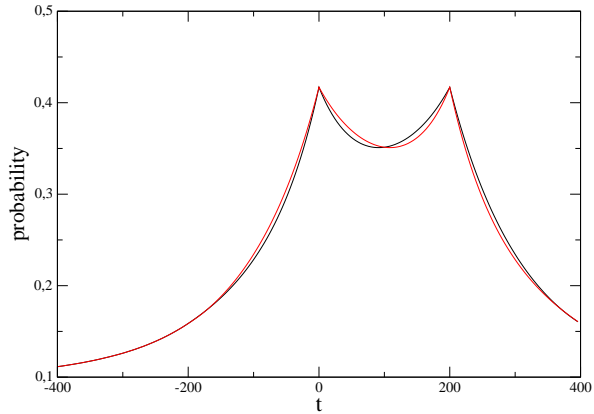}
 \caption{\label{Fig7} (Color on line) Probabilities $p(A,t-\tau;A,t\vert A)$ (black curve) and $p(A,-t;A,\tau-t\vert A)$ (red curve) for $\tau=200$ when the relation between the rates, Eq. (\ref{Eqgamma}), is not satisfied. This case corresponds to the arbitrary choice $\gamma_0=\gamma_2=0.003965$ and $\gamma_1=0.001516$. One can see that time-symmetry does not hold in this case.}
\end{center}
\end{figure}

The rest of the proof proceeds as in the case $N=1$. To illustrate the above demonstration, we show in Fig. \ref{Fig7} an example where Eq. (\ref{Eqgamma}) is violated and time-symmetry does not hold.

\section{Solution of the set of Eqs. (\ref{EqdiffAB})}

In this appendix we detail the solution of the set of linear differential equations  (\ref{EqdiffAB}). As in section III, all probabilities refer to the stationary state.

The expressions of the matrices ${\bf \Gamma}_A$ and ${\bf \Gamma}_B$ are
\begin{equation}
{\bf \Gamma}_A=
\begin{pmatrix}
0 & -\gamma_1 & -\gamma_1 & 0 & \gamma_1 & 0 & 0 & \gamma_1 & 0 & 0 \\
-\gamma_1 & -\gamma & 0 & -\gamma_1 & 0 & \gamma_0 & 0 & 0 & \gamma_2 & 0 \\
-\gamma_1 & 0 & -\gamma & -\gamma_1 & 0 & 0 & \gamma_2 & 0 & 0 & \gamma_0 \\
0 & -\gamma_1 & -\gamma_1 & 0 & \gamma_1 & 0 & 0 & \gamma_1 & 0 & 0 \\
\gamma_1 & 0 & 0 & \gamma_1 & \gamma & -\gamma_0 & -\gamma_2 & 0 & 0 & 0 \\
0 & 2\gamma_2 & 0 & 0 & -2\gamma_2 & 0 & 0 & 0 & 0 & 0 \\
0 & 0 & 2\gamma_0 & 0 & -2\gamma_0 & 0 & 0 & 0 & 0 & 0 \\
\gamma_1 & 0 & 0 & \gamma_1 & 0 & 0 & 0 & \gamma & -\gamma_2 & -\gamma_0 \\
0 & 2\gamma_0 & 0 & 0 & 0 & 0 & 0 & -2\gamma_0 & 0 & 0 \\
0 & 0 & 2\gamma_2 & 0 & 0 & 0 & 0 & -2\gamma_2 & 0 & 0
\end{pmatrix}
\end{equation}
and
\begin{equation}
{\bf \Gamma}_B=
\begin{pmatrix}
0 & -\gamma_2 & -\gamma_2 & 0 & \gamma_2 & 0 & 0 & \gamma_2 & 0 & 0 \\
-\gamma_0 & \gamma & 0 & -\gamma_2 & 0 & \gamma_1 & 0 & 0 & \gamma_1 & 0 \\
-\gamma_0 & 0 & \gamma & -\gamma_2 & 0 & 0 & \gamma_1 & 0 & 0 & \gamma_1 \\
0 & -\gamma_0 & -\gamma_0 & 0 & \gamma_0 & 0 & 0 & \gamma_0 & 0 & 0 \\
\gamma_0 & 0 & 0 & \gamma_2 & -\gamma & -\gamma_1 & -\gamma_1 & 0 & 0 & 0 \\
0 & 2\gamma_1 & 0 & 0 & -2\gamma_1 & 0 & 0 & 0 & 0 & 0 \\
0 & 0 & 2\gamma_1 & 0 & -2\gamma_1 & 0 & 0 & 0 & 0 & 0 \\
\gamma_0 & 0 & 0 & \gamma_2 & 0 & 0 & 0 & -\gamma & -\gamma_1 & -\gamma_1 \\
0 & 2\gamma_1 & 0 & 0 & 0 & 0 & 0 & -2\gamma_1 & 0 & 0 \\
0 & 0 & 2\gamma_1 & 0 & 0 & 0 & 0 & -2\gamma_1 & 0 & 0
\end{pmatrix}
\end{equation}
where $\gamma=\gamma_0+\gamma_2-2\gamma_1$.

These matrices have $4$ zero eigenvalues because the functions $p({\bf s}',t-\tau;{\bf s},t|{\bf s}^0)$ are not linearly independent. This is a consequence of the conservation of probabilities,
\begin{align}
\sum_{{\bf s},{\bf s}'} p({\bf s}',t-\tau;{\bf s},t|{\bf s}^0)&=1\nonumber\\
\sum_{{\bf s}',{\bf s}^0} p({\bf s}',t-\tau;{\bf s},t|{\bf s}^0)p({\bf s}^0)&=p({\bf s})\nonumber\\
\sum_{{\bf s},{\bf s}^0} p({\bf s}',t-\tau;{\bf s},t|{\bf s}^0)p({\bf s}^0)&=p({\bf s}') \ .
\end{align}
(Of course, these relations may be used to reduce the size of the matrices.)
The remaining $6$ eigenvalues of ${\bf \Gamma}_A$ and ${\bf \Gamma}_b$, denoted $\pm \lambda_{A,1},\pm \lambda_{A,2},\pm \lambda_{A,3},$ and $\pm \lambda_{B,1},\pm \lambda_{B,2},\pm \lambda_{B,3}$, respectively, are given by
\begin{align}
\lambda_{A,1}&=\sqrt{\gamma^2+8\gamma_1^2+8\gamma_0\gamma_2}\nonumber\\ 
\lambda_{A,2}&=\frac{1}{2}[\sqrt{\gamma^2+16\gamma_0\gamma_2}+\gamma]\nonumber\\
\lambda_{A,3}&=\frac{1}{2}[\sqrt{\gamma^2+16\gamma_0\gamma_2}-\gamma]
\end{align}
and
\begin{align}
\lambda_{B,1}&= \sqrt{\gamma^2+8\gamma_1^2+8\gamma_0\gamma_2}\nonumber\\
\lambda_{B,2}&= \frac{1}{2}[\sqrt{\gamma^2+16\gamma_1^2}+\gamma]\nonumber\\
\lambda_{B,3}&=\frac{1}{2}[\sqrt{\gamma^2+16\gamma_1^2}-\gamma] \ .
\end{align}
Therefore the two matrices have the same spectrum if the relation (\ref{Eqgamma}) $\gamma_0\gamma_2=\gamma_1^2$ is satisfied. The eigenvalues are then simply denoted $\pm \lambda_{1},\pm \lambda_{2},\pm \lambda_{3}$ with
\begin{align}
\lambda_{1}&= \sqrt{\gamma^2+16\gamma_1^2}\nonumber\\
\lambda_{2}&= \frac{1}{2}(\lambda_1+\gamma)\nonumber\\
\lambda_{3}&=\frac{1}{2}(\lambda_1-\gamma) \ ,
\end{align}
and Eqs. (\ref{Eqdiffp}) may be recast as
\begin{align}
{\bf p}_A(t)&={\bf U}_A^Te^{{\bf R}_At}({\bf U}_A^T)^{-1} {\bf p}_A(0)\nonumber\\
{\bf p}_B(t)&={\bf U}_B^Te^{{\bf R}_Bt}({\bf U}_B^T)^{-1} {\bf p}_B(0)
\end{align}
where ${\bf R}_A, {\bf R}_B$ are diagonal matrices with $(-\lambda_1,\lambda_1,-\lambda_2,\lambda_3,-\lambda_3,\lambda_2,0,0,0,0)$ and $(-\lambda_1,\lambda_1,-\lambda_3,\lambda_2,-\lambda_2,\lambda_3,0,0,0,0)$ on the diagonal, respectively, and the matrices ${\bf U}_A, {\bf U}_B$ read
\begin{equation}
{\bf U}_A=
\begin{pmatrix}
1&\lambda_2/2\gamma_1&\lambda_2/2\gamma_1&1&-\lambda_3/2\gamma_1&-\gamma_2/\gamma_1&-\gamma_0/\gamma_1&-\lambda_3/2\gamma_1&-\gamma_0/\gamma_1&-\gamma_2/\gamma_1\\
1&-\lambda_3/2\gamma_1&-\lambda_3/2\gamma_1&1&\lambda_2/2\gamma_1&-\gamma_2/\gamma_1&-\gamma_0/\gamma_1&\lambda_2/2\gamma_1&-\gamma_0/\gamma_1&-\gamma_2/\gamma_1\\
0&1&-1&0&0&-\lambda_3/2\gamma_0&\lambda_3/2\gamma_2&0&-\lambda_3/2\gamma_2&\lambda_3/2\gamma_0\\
0&1&-1&0&0&\lambda_2/2\gamma_0&-\lambda_2/2\gamma_2&0&\lambda_2/2\gamma_2&-\lambda_2/2\gamma_0\\
0&0&0&0&1&\lambda_2/2\gamma_0&\lambda_2/2\gamma_2&-1&-\lambda_2/2\gamma_2&-\lambda_2/2\gamma_0\\
0&0&0&0&1&-\lambda_3/2\gamma_0&-\lambda_3/2\gamma_2&-1&\lambda_3/2\gamma_2&\lambda_3/2\gamma_0\\
1&0&0&0&0&0&\gamma_1/\gamma_2&0&\gamma_1/\gamma_2&0\\
0&1&1&0&1&0&\gamma/\gamma_2&1&\gamma/\gamma_2&0\\
0&0&0&1&0&0&\gamma_1/\gamma_2&0&\gamma_1/\gamma_2&0\\
0&0&0&0&0&1&-\gamma_0/\gamma_2&0&-\gamma_0/\gamma_2&1
\end{pmatrix}
\end{equation}
\begin{equation}
{\bf U}_B=
\begin{pmatrix}
1&\lambda_3/2\gamma_2&\lambda_3/2\gamma_2&\gamma_0/\gamma_2&-\lambda_2/2\gamma_2&-\gamma_1/\gamma_2&-\gamma_1/\gamma_2&-\lambda_2/2\gamma_2&-\gamma_1/\gamma_2&-\gamma_1/\gamma_2\\
1&-\lambda_2/2\gamma_2&-\lambda_2/2\gamma_2&\gamma_0/\gamma_2&\lambda_3/2\gamma_2&-\gamma_1/\gamma_2&-\gamma_1/\gamma_2&\lambda_3/2\gamma_2&-\gamma_1/\gamma_2&-\gamma_1/\gamma_2\\
0&1&-1&0&0&-\lambda_2/2\gamma_1&\lambda_2/2\gamma_1&0&-\lambda_2/2\gamma_1&\lambda_2/2\gamma_1\\
0&1&-1&0&0&\lambda_3/2\gamma_1&-\lambda_3/2\gamma_1&0&\lambda_3/2\gamma_1&-\lambda_3/2\gamma_1\\
0&0&0&0&1&\lambda_3/2\gamma_1&\lambda_3/2\gamma_1&-1&-\lambda_3/2\gamma_1&-\lambda_3/2\gamma_1\\
0&0&0&0&1&-\lambda_2/2\gamma_1&-\lambda_2/2\gamma_1&-1&\lambda_2/2\gamma_1&\lambda_2/2\gamma_1\\
1&0&0&0&0&0&\gamma_0/\gamma_1&0&\gamma_0/\gamma_1&0\\
0&1&1&0&1&0&-\gamma/\gamma_1&1&-\gamma/\gamma_1&0\\
0&0&0&1&0&0&\gamma_2/\gamma_1&0&\gamma_2/\gamma_1&0\\
0&0&0&0&0&1&-1&0&-1&1 
\end{pmatrix} \ .
\end{equation}
The initial conditions are given by the vectors
\begin{equation}
{\bf p}_A(0)=\begin{pmatrix}
0 \\
0 \\
0 \\
0\\
p(B,\tau| A) \\
p(A,\tau | A) \\
p(D,\tau| A) \\
0 \\
0 \\
0\\
\end{pmatrix} , \
 {\bf p}_B(0)=\begin{pmatrix}
0\\
0\\
0 \\
0 \\
p(A,\tau| B) \\
p(B,\tau| B) \\
p(C,\tau| B) \\
0 \\
0 \\
0\\ 
\end{pmatrix} \ .
\end{equation}
Finally, solving the self-consistent equations (\ref{Eqselfcons}), we obtain
\begin{align}
\label{RES}
p(A,\tau| A)&= \frac{ \gamma_2}{K_A}[\lambda_1+\lambda_2e^{\lambda_3\tau}+\lambda_3e^{-\lambda_2\tau}] \nonumber\\
p(B,\tau| A)&= \frac{2\gamma_1^2}{K_A} [ e^{\lambda_3\tau}-e^{-\lambda_2\tau} ] \nonumber\\
p(D,\tau| A)&=\frac{\gamma_0}{K_A} [-\lambda_1+\lambda_2e^{\lambda_3\tau} +\lambda_3e^{-\lambda_2\tau} ] 
\end{align}
and
\begin{align}
p(A,\tau| B)&= \frac{2\gamma_1}{K_B} [1-e^{-\lambda_1\tau}]  \nonumber\\
p(B,\tau| B)&=\frac{1}{K_B}[\lambda_1e^{-\lambda_2\tau}+\lambda_2e^{-\lambda_1\tau}+\lambda_3 ] \nonumber\\
p(C,\tau| B)&=\frac{1}{K_B}[-\lambda_1e^{-\lambda_2\tau}+\lambda_2e^{-\lambda_1\tau}+\lambda_3 ] 
\end{align}
with 
\begin{align}
K_A&=\lambda_1(\gamma_2-\gamma_0)+[(\gamma_0+\gamma_2)\lambda_3-4\gamma_1^2]e^{-\lambda_2\tau}
+[(\gamma_0+\gamma_2)\lambda_2+4\gamma_1^2]e^{\lambda_3\tau}\nonumber\\
K_B&=2[\lambda_3+2\gamma_1+(\lambda_2-2\gamma_1)e^{-\lambda_1\tau}] \ .
\end{align}
As noted in section III, these nontrivial solutions of the self-consistent equations only exist when the rates satisfy Eq. (\ref{Eqgamma}). 

The stationary probability $p(A)$ is most easily computed by setting $t=\tau$ in the first equation (\ref{Eqprobflow}) which expresses the condition of detailed balance between the states A and B in the stationary state
 \begin{align}
 \label{EqpA}
 p(A)=\frac{1}{2}\: \frac{p(A,\tau\vert B)}{p(A,\tau\vert B)+p(B,\tau\vert A)} \  .
\end{align}
This yields
 \begin{align}
 \label{Finalresult}
p(A)=\frac{1}{2}\: \frac{K_A[1-e^{-\lambda_1\tau}]}{K_A[1-e^{-\lambda_1\tau}]+K_B\gamma_1[e^{\lambda_3\tau}-e^{-\lambda_2\tau}]} \ .
\end{align}
In particular, the values of $p(A)$ for $\tau=0$ and $\tau \rightarrow \infty$ are
\begin{align}
\label{tau0}
p_{0}(A)=\frac{\gamma_2}{2(\gamma_1+\gamma_2)}
\end{align}
and
\begin{align}
\label{tauinf}
p_{\infty}(A)=\frac{1}{2} \: \frac{4\gamma_1^2+\lambda_2(\gamma_0+\gamma_2)}{8\gamma_1^2+\lambda_2(\gamma_0+\gamma_2)+2\gamma_1\lambda_3} \ .
\end{align}


\begin{thebibliography}{10}

\bibitem{BPT1994} C. Van den Broeck, J. M. R. Parrondo, and R. Toral, Phys. Rev. Lett. {\bf 73}, 3395 (1994).
\bibitem{PRB2003} A. Pikovsky, M. Rosenblum, and J. Kurths, Synchronization, a Universal Concept in Nonlinear Sciences (Cambridge
University Press, Cambridge) (2001).
\bibitem{GHJM1998} L. Gammaitoni, P. H\"anggi, P. Jung, and F. Marchesoni, Rev. Mod. Phys. {\bf 70}, 223 (1998).
\bibitem{PK1997} A. Pikovsky and J. Kurths, Phys. Rev. Lett. {\bf 78}, 775 (1997).
\bibitem{YS1999} M. K. S. Yeung and S. H. Strogatz, Phys. Rev. Lett. {\bf 82}, 648 (1999).
\bibitem{AB} J. A. D. Appleby and E. Buckwar, Dynam. Systems and Appl. {\bf 14}, 175 (2003).
\bibitem{BVTH2005} D. Bratsun, D. Volfson, L. S. Tsimring, and J. Hasty, Proc. Natl. Acad. Sci. USA {\bf 102},14593 (2005).
\bibitem{CWZA2005}  L. Chen, R. Wang, T. Zhou, and K. Aihara, Bioinformatics  {\bf 21}, 2722 (2005).   
\bibitem{GLL1999} S. Guillouzic, I. L'Heureux, and A. Longtin, Phys. Rev. E {\bf 59}, 3970 (1999); Phys. Rev. E {\bf 61}, 4906 (2000).
\bibitem{F2002} T. D. Frank, Phys. Rev. E {\bf 66}, 011914 (2002).
\bibitem{KM1992} U. K\"uchler and B. Mensch, Stoch. Rep. {\bf 40}, 23 (1992).
\bibitem{FB2001} T. D. Frank, and P. J. Beek, Phys. Rev. E {\bf 64}, 021917 (2001); T. D. Frank, P. J. Beek, and R. Friedrich, Phys. Rev. E {\bf 68}, 021912 (2003).
\bibitem{ZXXL2009} H. Zhang, W. Xu, Y. Xu, and D. Li, Physica A {\bf 388},3017 (2009).
\bibitem{F2005} T. D. Frank, Phys. Rev. E {\bf 71}, 031106 (2005).
\bibitem{HGBMH2004} J. Houlihan, D. Goulding, T. Busch, C. Masoller, and G.
Huyet, Phys. Rev. Lett. {\bf 92}, 050601 (2004).
\bibitem{GMTG2007} G. M. Gonz\'alez, C. Masoller, M. C. Torrent, and J. Garc\'ia Ojalvo, Eur. Phys. Lett. {\bf 79}, 64003 (2007).
\bibitem{DZ1978} R. C. Desai and R. Zwanzig, J. Stat. Phys. {\bf 19}, 1 (1978).
\bibitem{JBPM1992} P. Jung, U. Behn, E. Pantazelou, and F. Moss, Phys. Rev. A {\bf 46}, R1709 (1992).
\bibitem{TP2001} L. S. Tsimring and A. Pikovsky, Phys. Rev. Lett. {\bf 87}, 250602 (2001).
\bibitem{M2003} C. Masoller, Phys. Rev. Lett. {\bf 90}, 020601 (2003).
\bibitem{HT2003} D. Huber and L. S. Tsimring, Phys. Rev. Lett. {\bf 91}, 260601 (2003); Phys. Rev. E  {\bf 71},
   036150 (2005).
\bibitem{G2009} T. Galla, Phys. Rev. E  {\bf 80}, 021909 (2009).
\bibitem{KM2009} M. Kimizuka and T. Munakata, Phys. Rev. E {\bf 80}, 021139 (2009). 
\bibitem{K1940} H. Kramers, Physica (Utrecht) {\bf 7}, 284 (1940).
\bibitem{PZC2002} A. Pikovsky, A. Zaikin, and M. A. de la Casa, Phys. Rev. Lett. {\bf 88}, 050601 (2002).


\end{thebibliography}
\end{document}